\documentclass[12pt]{article}
\input{psfig}

\makeatletter
\@addtoreset{equation}{section}

\begin{document}

\newcommand{\be}{\begin{equation}} \newcommand{\ee}{\end{equation}}
\newcommand{\ba}{\begin{eqnarray}} \newcommand{\ea}{\end{eqnarray}}
\newcommand{\qq}{Q {\bar Q}}       \newcommand{\la}{\lambda}
\newcommand{\N}{{\cal N}}              
\newcommand{\W}{{\cal W}}
\newcommand{\llc}{\langle  \tr \la \la \rangle} 
\newcommand{\wg}{\wedge}
\newcommand{\ra}{\rightarrow}
\newcommand{\apt}{{\tilde \alpha } '}
\newcommand{\tr}{{\mbox tr}}
\newcommand{\T}{{\cal T}}
\newcommand{\srt}{\sqrt{2}}

\begin{titlepage}
\rightline{hep-th/0103163}
\rightline{TAUP-2658-01}
\vskip 1cm
\centerline{{\Large \bf On the holographic duals of $\N=1$ gauge dynamics}}

\vskip 1cm
\centerline{A. Loewy and J. Sonnenschein}

\vskip 1cm
\begin{center}
School of Physics and Astronomy,
\\Beverly and Raymond Sackler Faculty of Exact Sciences,
\\Tel Aviv University, Ramat Aviv, 69978, Israel.
\end{center}
\vskip 1cm

We analyze the holographic description of 
several properties of $\N=1$ confining gauge dynamics. 
In particular we discuss  
Wilson loops including the issues of a L\"uscher term  and the broadening
of the flux tubes, 't Hooft loops, baryons, instantons, chiral
symmetry breaking, the gluino condensate and  BPS domain walls.

\end{titlepage}

\section{Introduction}

On the route from the original AdS/CFT  duality \cite{Mald,GKP,W3} to
a holographic
 description of realistic  gauge  dynamics, several approaches were used to dualize 
field theories with $\N=1$ supersymmetry \cite{PS,KW,KT,RT,KS,MN1,MN}.
Up to date the holographic description of $\N=1$ SYM theory has not
been written down. However, the models due to Klebanov and Strassler (KS) \cite{KS} and Maldacena
 and Nunez (MN) \cite{MN} made an important step towards this goal.

We will first  examine the  
features of $\N=1$ gauge theories that can be reliably computed 
from the  holographic dual models. This will be  done  mainly in the
context of the KS and MN models.   
Among the gauge theory properties that we address are 
Wilson loops including the issue of the L\"uscher term, the broadening
of the flux tube, 't Hooft loops, baryons, instantons, chiral symmetry breaking, the gluino condensate, and BPS domain walls.
In addition we describe several other possible probes made out of certain  
brane configurations for which there are no corresponding 4d gauge theory
states.

The second task of this work is to extract from these properties
certain guide lines for the construction of other supergravity duals of $\N=1$ confining theories. 
Based on the very few models in the market, we cannot come up
with a precise recipe based on a set of  restrictive rules, 
but rather only with certain ``recommendations'' for the model builder. 
An important role in these guide lines is played by wrapped branes.
Wrapping of Euclidean Dp-brane over $p+1$ cycles, Dp-branes over $p$ cycles,  
Dp-branes over $p-1$ cycles and Dp-branes  over $p-2$ cycles are argued to 
correspond to instantons, baryons, 't Hooft loops and BPS domain walls. 
There are also string theory descriptions of $\N=1$ SYM using brane configurations \cite{GivKut}, MQCD \cite{W2}, and brane engineering \cite{KV}. In this paper we will concentrate only on holography. 

The paper is organized as follows. In section 2 we briefly summarize the
supergravity backgrounds proposed by KS and MN.
Wilson loops are discussed in section 3.  The area law behavior of the two
models is extracted from the classical supergravity. We then argue
that there  exists an attractive
L\"uscher term, broadening without a roughening phase transition and
Regge-like trajectories. 
In section 4 we discuss baryon configurations in a general confining
theory and in the KS and MN models in particular. 
Section 5 is devoted to the breaking of the $U(1)_R$ symmetry
 to $Z_{2N}$ via instantons and the spontaneous breaking down to $Z_2$.
The description of the gluino condensate in terms of the background
3-form is presented in section 6. 
In section 7 the supergravity configurations that correspond to BPS
domain walls are discussed and their tension is computed. Additional
brane probes are discussed in section 8. In section 9 we discuss the constraints that each of the gauge dynamics properties imposes on supergravity backgrounds that  correspond to    
 $\N=1$ gauge theories.

\section{Brief review of the  KS and MN models}
 
The original AdS/CFT correspondence can be generalized to $\N=1$ theories in many ways. One way to explicitly break some of the supersymmetries is to place the D3-branes at a conifold singularity \cite{KW}. The world-volume gauge theory on the $N$  D3-branes is an $SU(N) \times SU(N)$ $\N=1$ superconformal gauge theory with bi-fundamental chiral multiplets $A_i$ and $B_i$ $(i=1,2)$ in the $({\bar N},N)$ and $(N,{\bar N})$ representations of the gauge group. These chiral multiplets transform under the global $SU(2) \times SU(2) \times U(1)$ symmetry as $(2,1)_1$ and $(1,2)_{-1}$. The superpotential is given by 
\be \W \sim \tr (A_1 B_1A_2 B_2 -A_1 B_2 A_2 B_1)  \ee
The supergravity dual of this theory is an $AdS_5 \times T^{1,1}$ geometry. The correspondence in this background has been worked out in \cite{Gub,CDDF}. 

It is also possible to add $M$ fractional D3-branes, which are D5-branes wrapped on an $S^2$ of the conifold base \cite{KT}. The world-volume theory in such a case is $SU(N+M) \times SU(N)$. The addition of $M$ fractional branes explicitly breaks the conformal symmetry. The naive supergravity dual of this theory, found by Klebanov and Tseytlin (KT) \cite{KT},  has a naked singularity at the origin, and there are non-trivial $F_5$ and $G_3=F_3 + i H_3$ profiles related to the number of regular and fractional D3-branes, the dilaton stays constant.
$$ \alpha'^{-1} ds^2 = \frac{u^2}{g_sM \sqrt{\ln(u/u_0)}} dx_{0123}^2 + \frac{g_sM \sqrt{\ln(u/u_0)}}{u^2} du^2 + g_s M \sqrt{\ln(u/u_0)} ds_{T^{1,1}}^2, $$
$$ B_2 = \frac{3}{2}g_sM (\sin \theta_1 d\theta_1 \wg d\phi_1 -\sin \theta_2 d\theta_2 \wg d\phi_2) \ln(u/u_0), $$   
\be \int_{S^{3}} F_3 = M, \qquad \int_{T^{1,1}} F_5 = N + g_s M^2 \ln(u/u_0)\ee
The two gauge couplings are scale dependent 
\be \label{rgcks} \frac{1}{g_1^2}-\frac{1}{g_2^2} \sim e^{-\phi} \Big[ \Big( \int_{S^2} B_2 \Big) - \frac{1}{2} \Big]  \sim M \ln (u/u_0), \quad \frac{1}{g_1^2}+\frac{1}{g_2^2} \sim e^{-\phi} = {\mbox const.} \ee
This reproduces the logarithmic running of the gauge couplings expected in $\N=1$ gauge theories.

Klebanov and Strassler \cite{KS} have proposed that the naked singularity can be resolved by replacing the conifold with a deformed conifold in which the $S^3$ part of the conifold base stays at a finite radius near the origin. The logarithmic decrease in the 5-form flux as we flow to the IR was interpreted as a cascade of Seiberg dualities that occur each time one of the coupling constants diverges. In the following we will assume that $M$ is such that at the bottom of the duality cascade we are left with an $SU(M)$ gauge theory. However, this is not a holographic dual of pure SYM, since for $g_s M \gg 1$, where supergravity is a valid approximation, the duality cascade is dense. In other words, there is no finite energy range in which the theory is pure SYM.

The supergravity solution of the deformed conifold is of the following form:
\ba \label{ksb} ds^2 = h^{-1/2}(\tau) dx_{0123}^2 + h^{1/2}(\tau) ds^2_{6}, \ea
where $ds^2_{6}$ is the metric of the deformed conifold 
\be \label{defcon} ds^2_{6} = \frac{1}{2} \epsilon^{4/3}
K(\tau) \Big[ \frac{1}{3 K^3 (\tau)} (d\tau^2 + (g^5)^2)  +
 \cosh^2  \Big(\frac{\tau}{2}\Big) [(g^3)^2 + (g^4)^2] + \sinh^2 \Big(\frac{\tau}{2}\Big) [(g^1)^2 + (g^2)^2 ] \Big]. \ee
The $g^i$'s are a set of 1-forms parameterizing the conifold
$$\srt g_1=-\sin\theta_1 d\phi_1-\cos\psi\sin\theta_2 d\phi_2+\sin\psi d\theta_2, \quad \srt g_2=d\theta_1-\sin\psi\sin\theta_2 d\phi_2-\cos\psi d\theta_2, $$
$$\srt g_3=-\sin\theta_1 d\phi_1+\cos\psi\sin\theta_2 d\phi_2-\sin\psi d\theta_2, \quad \srt g_4=d\theta_1+\sin\psi\sin\theta_2 d\phi_2+\cos\psi d\theta_2, $$
\be \label{gi} g_5=d\psi+\cos\theta_1 d\phi_1+\cos\theta_2 d\phi_2,\quad K(\tau)= \frac{(\sinh(2\tau)-2\tau)^{1/3}}{2^{1/3}\sinh \tau}.\ee
The function $h(\tau)$ is given by 
\be  \label{h(t)} h(\tau) \sim  \alpha \int^{\infty}_{\tau} dx \frac{x \coth x -1}{\sinh^2 x} (\sinh(2x) -2x)^{1/3}. \ee
The asymptotic form of this function is $h( \tau \ra \infty) \ra \alpha \tau
e^{-4u/3}$. For large $\tau$ we can make a change of variables $u^3 \sim
\epsilon^2 e^{\tau} \sim m^3 e^{\tau}$. This brings $h(u)$ to the familiar form 
$h(u) \ra \alpha \frac{\ln(u / \epsilon^{2/3})}{u^4}$      
which is the KT solution, with $\alpha \sim (g_s M)^2$. $F_3$ and $B_2$ are given by 
\be F_3 = M \Big[ g^5 \wg g^3 \wg g^4 + d[F(\tau)(g^1 \wg g^3 + g^2 \wg g^4)] \Big], \ee
\be B_2 = g_s M \Big[f_-(\tau) g^1 \wg g^2 + f_+(\tau) g^3 \wg g^4 \Big] , \ee
where 
\be F(\tau)=\frac{\sinh \tau - \tau}{2 \sinh \tau}, \qquad f_{\pm}(\tau)=\frac{\tau \coth \tau -1}{2 \sinh \tau}(\cosh \tau \pm 1). \ee

Another possible supergravity dual of pure SYM theory was proposed by Maldacena and Nunez \cite{MN}. They considered $N$ NS5-branes wrapped on an $S^2$ in the context of 7d gauged supergravity, with boundary conditions that preserve 4 supersymmetries. Their solution is directly related to a non-Abelian monopole solution found in \cite{CV}. The 10d background is:
\be \label{mnbns} ds^2 =  dx_{0123}^2 + \apt N \Big[ d\tau^2 +
e^{2g(\tau)} (e_1^2 + e_2^2 ) + \frac{1}{4} ( e_3^2 + e_4^2 + e_5^2 )\Big], \ee
\be e^{2 \phi} = e^{2\phi_0} \frac{2 e^{g(\tau)}}{\sinh 2\tau}, \qquad  e^{2 g(\tau)} = \tau \coth 2\tau - \frac{\tau^2}{\sinh^2 2\tau} - \frac{1}{4}. \ee
The $e_i$'s are a set of 1-form given by
$$e_1= d \theta_1, \quad e_2 = \sin \theta_1 d \phi_1, $$
$$e_3= \cos \psi d \theta_2 + \sin \psi \sin \theta_2 d \phi_2 - a(\tau) d \theta_1, $$
$$e_4= - \sin \psi d \theta_2 + \cos \psi \sin \theta_2 d \phi_2 - a(\tau) \sin \theta_1 d \phi_1, $$
\be e_5= d \psi + \cos \theta_2 d \phi_2 - \cos \theta_1 d \phi_1, \quad a(\tau) = \frac{\tau^2}{\sinh^2 \tau}. \ee 
The 4d gauge coupling is related to the 6d gauge coupling by
\be \frac{1}{g_{(4)}^2} = \frac{\mbox{vol}(S^2)}{g_{(6)}^2} = \frac{N e^{2g}}{2 \pi^2},\ee
where $g_{(6)}^2=(2\pi)^3\apt$, so that the 4d coupling is dimensionless. From the asymptotic form of the function $e^{2g(\tau)} \ra \tau$, and a change of variables to $\tau\sim\ln(u/m)$, we can see that the 4d gauge coupling runs logarithmically.
When flowing to the IR the string coupling becomes of order one, and we must use the S-dual description. 
\be \label{mnbd5} ds^2 =  e^{\phi} \Big[dx_{0123}^2 + \alpha' g_s N (d\tau^2 +
e^{2g(\tau)} (e_1^2 + e_2^2) + \frac{1}{4} (e_3^2 + e_4^2 + e_5^2)) \Big], \quad e^{2 \phi} = e^{-2\phi_0} \frac{\sinh 2\tau}{2 e^{g(\tau)}}, \ee
with $H_3$ replaced by $F_3$. Note that we keep factors of $\alpha' g_s$ explicit in (\ref{mnbd5}), and take $e^{-2\phi_0} \sim g_sN$. 

As in the case of the KS background only in the extreme IR is this
background supposed to be dual to pure SYM. At higher energies there will be KK states from $S^3$. In order to decouple these KK states one would like to take the radius of the 3-sphere to zero, $\sqrt{g_s N} \ra 0$, but that will produce a region of large curvature, since in both models $\alpha' {\cal R} \sim  1/\sqrt{g_s N}$. 
Both backgrounds are special cases of branes wrapping manifolds with a
$R \times S^2 \times S^3$ topology \cite{TP,TPZ}. In the case of $N$
D5-branes and no D3-branes we get the MN solution, and for M D5-branes
and N D3-branes with $N$ a multiple of $M$ we get the KS solution.


\section{Wilson loops, flux tubes and 't Hooft loops} 

The long distance quark anti-quark potential is one quantity that is likely to
be  invariant to the exact dynamics in the UV.  From a holographic point of view these backgrounds differ from other supergravity duals of confining theories in that the metric does not have a horizon at a finite radius \cite{W1}. The area law in these cases is a consequence of the fact that $f^2(\tau)=g_{tt}g_{xx}(\tau)$ has a minimum at $\tau=0$, so a fundamental string ``prefers'' to be on the hyper-surface $\tau=0$. 
The finite string tension in both the MN and
KS  backgrounds is proportional to $f(0)$, and is $\tau_s= \sqrt{g_sN}/2\pi \alpha'$ in the MN model and $\tau_s =m^2/g_sM$ in the KS model. 
This form of confinement was previously observed in MQCD calculations \cite{W2,HOO,HS,KSS0}. Although MQCD is not a holographic description, the Wilson loop can be evaluated in much the same way as in the AdS/CFT. 

The action of a fundamental string in the above backgrounds is of the
usual form 
\be \label{act} S = \int dx dt \sqrt{f^2(\tau) + g^2(\tau) \tau'^2}, \ee
where $g^2(\tau)=g_{tt}g_{\tau \tau}$.
It is easy to see from the metrics in the introduction that we have
\be  f^2(\tau)=h^{-1}(\tau), \qquad g^2(\tau)=1, \qquad \mbox{deformed conifold}. \ee
\be f^2(\tau)  = g_sN  \frac{\sinh
  2 \tau}{2e^{g(\tau)}}, \qquad g^2(\tau) = \alpha' g_s N  f^2(\tau) \qquad \mbox{wrapped D5}. \ee

In calculating the action of the Wilson loop one has to add also  the coupling of the string
to the dilaton and to $B_{NS}$. 
 In  the KS model of the dilaton is constant and in
MN background there is a non-trivial dilaton.  However, in the case of an 
infinite-strip Wilson loop the Gaussian world-sheet curvature vanishes so there is no coupling to the dilaton at leading order in $g_s$. 
 Both the MN and KS backgrounds have  non-trivial $B_{NS}$ profiles.
The only components of $B_{NS}$ that can couple to the world-sheet are $B_{01}, B_{0\tau}, B_{1\tau}$. For the backgrounds in question they vanish.
Thus, for the two models we investigate here 
the two additional  couplings do not contribute in the leading order
in $g_s$, and the action of the string includes only the Nambu-Goto term (\ref{act}).

We can use the general theorems in \cite{KSS1} to calculate the
classical corrections to the linear quark anti-quark potential. The
key observation is that in both cases $f(\tau)$ can be approximated
near $\tau=0$ by $f(\tau)=a_0 + a_2 \tau^2 +O(\tau^4)$ and $g(\tau)=b_0+O(\tau^2)$. 
The resulting potential is therefore
\be \label{potential} E = f(0)L -2 \kappa + O((\log (L))^{\beta}
e^{-\alpha L}).\ee
The parameters in the potential are \cite{KSS1}
\be \kappa=\int^{\infty}_{0} \frac{g(\tau)}{f(\tau)} 
\Big( f(\tau)- \sqrt{f^2(\tau)-f^2(0)} \Big) d\tau, \qquad \alpha=\frac{\sqrt{2 f(0) a_2}}{b_0}. \ee
Since in both cases $\alpha \neq 0$, the classical correction to the linear potential are exponentially small. This was also the case in other confining backgrounds. 

Stringy corrections in the AdS/CFT can come from two sources. One
possible source is a correction to the background metric and
fields. Since the backgrounds reviewed in the previous section are not group manifolds or coset spaces like $AdS_5 \times S^5$ we really do not have any control on these corrections. Such stringy corrections will change the string tension, but keep the leading linear $L$ dependence. Another source of stringy corrections that becomes relevant in Wilson loop calculations is the fluctuating world-sheet \cite{KSS3,DGT,FGT,GO1}. Since the classical configuration is a flat horizontal string, these corrections will lead to a quark anti-quark potential with a L\"uscher term correction of the form 
\be V = f(0) L - \frac{\pi (n_b - n_f)}{24} \frac{1}{L}, \ee
where $n_b$ and $n_f$ are the number of massless bosonic and fermionic
excitations of the string. Close to $\tau=0$ the metric is
effectively $R^{1,6} \times S^3$ . The $S^3$ stays at a finite radius proportional to $\sqrt{g_s N}$. The number of massless bosonic excitations will thus be
$n_b=8$ (or 7 according to \cite{KSS3}) if the distance between the quarks is smaller than the radius of
$S^3$, and $n_b=5$ if it is larger.  

Next we consider the fermionic fluctuations. In \cite{KSS3} it was argued
that in the case of the $AdS_5$ black hole background, which for very high
temperature is dual
to a 3d Yang-Mills theory, the fermionic modes of a horizontal string
are all massive. This is due to the coupling of the fermionic
modes to the RR $F_5$. The KS background also has a RR $F_5$, but near $\tau=0$ it vanishes. However, there is a non-vanishing $G_3$, which for $\tau=0$ is in the $S^3$ direction. 
Since we do not have a candidate string action in this background, we
cannot be certain whether the fermion excitations have zero modes or are massive in this background. 
If the coupling of fermions to the RR 3-form is of the same nature as the
coupling to $F_5$, then the fermionic determinant will be that of massive modes. However, we do have a field theory argument that there are massless fermions. The dual theory in the IR, where this analysis is valid, is pure SYM  plus massive KK states. SYM does not have BPS saturated
strings. 
Therefore, we expect the string to have 2 left and 2 right
massless fermionic excitations reflecting the fact that it is not
invariant under any of the 4 supersymmetries \cite{W2}.
To summarize, if we combine our arguments about the bosonic and fermionic modes it seems that there is an attractive L\"uscher term in the $\N=1$ models under review, after all even in the $\N=4$ case one believes that this term does not vanish \cite{FGT,DGT,KSS3}, however at present we cannot show it explicitly. 

It was noted that the confining backgrounds such as \cite{W1}
reproduce some field theory phenomena such as broadening of the flux tube \cite{GO} and 
Regge like trajectories \cite{Jan}. We will now argue that in the framework of holographic duals these phenomena are generic to a large class of confining
backgrounds, and in particular to the two under discussion.
The fact that the flux tube broadens already at strong coupling means
that one can hope to extrapolate strong coupling results to weak
coupling without running into a phase transition. In \cite{LMW} it was
shown that a stringy model of a flux tube in flat space leads to a
logarithmic broadening of the flux tube as a function of the distance 
between the quarks. It was shown in \cite{GO} that this is also the case
in the $AdS_5$ black hole metric. The way to calculate this effect is 
to consider the surface between two concentric Wilson loops with radii
$R_1 , R_2$ a distance $H$ apart along the $z$-axis. The area of the surface is given by
\be S=\frac{1}{2 \pi \alpha'} \int r dr \sqrt{f^2(\tau)(1+z'^{2})+g^2(\tau) \tau'^{2}} \ee
Substituting the equation of motion for $z'$ into $S$ we get
\be S=\frac{1}{2 \pi \alpha'}\int r dr \sqrt{F^2(\tau) + G^2(\tau) \tau'^2} \ee
\be F^2=\frac{r^2f^4}{r^2f^2-q^2}, \qquad G^2=\frac{r^2f^2g^2}{r^2f^2-q^2}, \ee
where $q$ is an integration constant. The qualitative picture that arises is that if the new $F$ and $G$
lead to confinement then for $R_1 \gg H \gg R_2$ the surface will
concentrate at some $\tau=\tau_0$, and there the flat case analysis is
valid, and leads to broadening. The two possible scenarios for
confinement are: (i) $g(\tau)$ diverges at some $\tau_{div}$ while $f(\tau_{div})$ stays
finite. In this case it is clear that $G(\tau)$ and $F(\tau)$ will also have
these properties. (ii) $f(\tau)$ has a minimum  at $\tau_{min}$.
 In this case $F(\tau)$ will
also have a minimum, although not necessarily at the same $\tau$. We
conclude that every confining background will lead to flux tube
broadening.

The same reasoning applies to recent work by Janik \cite{Jan} that uses
holography to describe the Pomeron. The results in that case are
classically the same as would be expected from a dual string model in flat
4 dimensions. It is only when one considers the stringy corrections to
the intercept that the fact that the string is embedded in more than 4
dimensions comes into play. Therefore it seems likely that all
confining backgrounds will produce a Regge trajectory.

The ``dual'' phenomenon of quark confinement is the screening of magnetic monopoles.
The later is observed via  the 't Hooft loop. 
Supergravity  duals of 't Hooft loops were constructed for models of confining quarks \cite{BISY,PS}.
In those cases one uses un-wrapped D1-branes, or D2-branes in type IIA \cite{GrosOo,Li}. However, one can easily check that these objects are not adequate in the MN and KS models, since a D1-brane will have a non-vanishing string tension. It is not clear to us what field theory object it corresponds to in the 4d theory as we further discuss in section 8.

Instead  it was argued \cite{KS,MN} that the holographic 
description of  gauge theory monopoles in both backgrounds 
are fractional D1-branes. As pointed out in \cite{KS} one can only prove that they are not confined. The action of a D3-brane wrapped on an $S^2$ will have the same general form as the action of the string, but with the functions $f^2(\tau)$ and $g^2(\tau)$ multiplied by the volume of the $S^2$. In both cases this volume vanishes at $\tau=0$ so the monopoles have vanishing string tension and are not confined.  

\section{Baryons}

The  baryon vertex of $\N=4$  $SU(N)$ SYM is dual to a
D5-brane wrapping $S^5$ in $AdS_5\times S^5$. This vertex is connected to $N$ 
points on the boundary of the $AdS_5$ with $N$ fundamental strings \cite{Wittenbar}.
The analysis of a configuration of this type  in dual supergravity models of
confining gauge theories \cite{BISY,CGST} 
 showed their stability and led to a computation of the baryonic
mass.  

In the holographic models of $\N=1$  there are two types of ``building blocks''
for  the  construction of the baryonic vertex:    
(i)   wrapped D3-brane over
$S^3$ and (ii)   D5-branes that wrap the compact 5 dimensional manifold,
for instance the base of the deformed conifold in the KS model.
There are two types of baryons to which one needs to provide supergravity duals. The first type is similar to the original  $AdS_5\times S^5$  baryon, namely
a baryon constructed from external quarks. The second type is  a
baryon built from the ``elementary particles'' of the theory. 

Prior to the discussion of the 
assignment of wrapped branes to the two different  types of baryons, we
first follow the simplified method 
\footnote{An improved
   approximation  which also incorporates the gauge field
   on the brane was used in \cite{CGST}. This type of calculation
   can also be applied for the present case, however to deduce the basic
properties of the system the simplified approach suffices}
 of \cite{BISY} for a general background with $N$ units of flux.
The basic configuration is that of a baryonic vertex located at a radial
position $\tau_0$, connected with $N$ strings to $N$ ``quarks''
distributed in a symmetric way on a circle of radius $L$ on the
boundary. The baryonic vertex is taken to be a wrapped Dp-brane around some p-cycle at $\tau_0$.
  
Let $S_{Dp}(\tau_0)$ denote the action of the wrapped Dp-brane at $\tau_0$. It has the following form
\be
S_{Dp}(\tau_0)= {N\over (2\pi)^p \alpha'^{(p+1)/2}} \int
d^{p} \theta e^{-\phi} \sqrt{\det \  g(\tau_0)},
\ee
where $g$ is the induced metric on the wrapped brane, and the $\theta$'s are
coordinates on the p-cycle the brane wraps.
The $N$ strings attached to this vertex have a Nambu-Goto action. 
The total action is  $S_{Dp} + NS_{F1}$.
The difference between the baryon and Wilson loop calculations is 
that in the baryon case the strings have one end not on the boundary
but rather on the baryon vertex. This produces a surface term in the 
string action, which must be equal to the force the baryon 
vertex  applies on the strings for the whole system to be static. 
\be N \frac{g^2(\tau_0) \tau_0' \delta \tau_0}{\sqrt{f^2(\tau_0) 
+ g^2(\tau_0) (\tau_0')^2}} = \partial_{\tau_0}S_{Dp}(\tau_0) \delta
\tau_0, \ee
where the $\tau_0'$ denotes the value of $\frac{d\tau}{dx}$ at
$\tau_0$. 
 From this condition one can get the radius
 and energy of the baryon by the same procedure as for the quark anti-quark potential
$$ L = \int_{\tau_0}^{\infty} d\tau \frac{g(\tau)}{f(\tau)}
\frac{1}{\sqrt{\beta f^2(\tau) -1}}, \qquad \beta = 
\frac{g^2(\tau_0) / f^2(\tau_0)} {g^2(\tau_0) - (\partial_{\tau_0}S_{Dp}(\tau_0))^2}, $$
\be E= N \int_{\tau_0}^{\infty} d\tau
\frac{\sqrt{\beta}f(\tau)g(\tau)}{\sqrt{\beta f^2(\tau) -1}} - N
\int_0^{\infty} d\tau g(\tau). \ee
In confining backgrounds both integrals receive most of their contributions from the region $\beta f^2(\tau)\approx 1$, and $f(\tau)$, $g(\tau)$ in the KS
and MN backgrounds are regular functions with a minimum at $\tau=0$ so
the energy of the baryon is linearly proportional to its size $L$ and to $N$. 

Now we would like to check which wrapped branes correspond to what baryons.
In the conformal case \cite{KW} there are two types of baryons built from composites of $N$ $A$ particles and $N$ $B$ particles \cite{GK}. 
The corresponding brane configurations involves wrapping a D3-brane  over two supersymmetric cycles of minimal volume associated with constant $(\theta_1,\phi_1)$
or $(\theta_2, \phi_2)$. 
The dimensions of these operators were computed and were shown to 
be identical to the field theory dimensions.
In a similar manner one can write down two type of baryonic operators
in  the  KS model \cite{Aharony}.
In the UV the baryons  are  built from  a composite of ${N+M\over M}$ (for $N\ mod \ M =0$)
 singlets of the $SU(N)$ group factors  which
themselves are composites of $N$ $A$ particles (or $B$ particles).
Replacing $N$ with $P=N-nM$ in this construction produces the baryons 
after $n$ steps of the Seiberg duality. At the end of the cascade,
where $P=0$, there are no matter fields in the left over theory with $SU(M)$
gauge group and
hence there are no baryons composed from ``dynamical  quarks'', but
one can still construct a baryon made out of ``external quarks''.
 
In \cite{Aharony} a proposal for  the dual configuration was suggested 
based on the  following configuration. In the  asymptotic regime
it is a set of $M+N \over M$ D3-branes wrapped on the $S^3$ with $M$ string ending on each
D3-brane in such a way that the total of $N+M$ strings end on their other side
on a single D5-brane that wraps the five-cycle. 
The number of strings attached to any D3-brane and the D5-brane is
obviously the amount of the $G_3$ and $F_5$ flux.
This constitutes the required baryon operators in the 
region where the flux of $F_5$ is  $N+M$ and that of $G_3$ is $M$.
 In the region where the former is $p+M$ and the latter is still $M$, a
 similar construction holds with $M+P\over M$ wrapped D3-branes
 connected to a single wrapped D5-brane with $M+P$ strings.
Finally in the IR region where $p=0$ the $M$ strings stretching out of
the wrapped D3-brane cannot be attached to the 5-brane since   now
the $F_5$ flux vanishes and hence can end only on $M$ external quarks. 

\section {Instantons, and the breaking of $U(1)_R$ to $Z_{2N}$} 

Instantons in $\N=1$ SYM theory break the $U(1)_R$ symmetry to $Z_{2N}$. 
To exactly identify the dual of the field theory instanton 
we need an object with an action of the field theory instanton, 
\be \label{inst} S_{inst.} = \frac{8 \pi^2}{g_{YM}^2(\mu)} + i \theta_{FT}. \ee
In the large $N$ limit
the breaking of the $U(1)_R$ to $Z_{2N}$ cannot be seen in the 
isometries of the classical background. In order to see this breaking one has to consider instanton probes, and to show that their action is invariant only under the $Z_{2N}$ symmetry.
The supergravity dual of the 
field theory instanton can in general be a combination of a D(-1) and a wrapped world-sheet of a D1/F1-string. 
 
In the MN background the $SU(N)$ instanton was argued to be a D1-brane wrapping the $S^2$ 
and another $S^2$ inside the $S^3$ which is defined by
 $\theta=\theta_1=\theta_2$ and $\phi=\phi_1=\phi_2$ \cite{MN}. 
In the UV one uses the S-dual background and the D1-brane is replaced with a fundamental string. Let us see what is the action of such a string. For simplicity we will carry out all calculations in the UV, and use the singular MN metric. The value of the induced $B_{\theta \phi}$ on the specified 2-cycle is $B_{\theta \phi} = -\frac{N \psi}{2} \sin \theta$. Therefore the imaginary part of the world-sheet action of the F1-strings comes from the WZ term 
\be {1\over 2\pi}\int_\psi d \theta d \phi B_{\theta \phi}= b-N \psi. \ee
where $b$ is some integration constant. 
This flux should be identified with the phase $\theta_{FT}$ in the field theory action so it is clear that only $Z_{2N}$ rotations of this phase $\psi \ra \psi +{2\pi k \over N}$ leave the path integral invariant. 

Now let us compute the real part of the action, $S_R$ of this configuration. By wrapping the string world-sheet over the same cycle we get a Nambu-Goto action of the form
\be \label{mninst} S_R = \frac{1}{2 \pi \apt} \int d\theta d\phi \sqrt{g} = 2N \sqrt{e^{4g(\tau)} + \frac{1}{2} e^{2g(\tau)}+\frac{1}{16}} \ee
In the UV, where $e^{2g(\tau)} \sim \tau$, the first term is dominant and we get $S_R = 2 N e^{2g(\tau)} = \frac{4 \pi^2}{g_{YM}^2}$.
The instanton action in the UV is indeed proportional to $g_{YM}^{-2}$, but there are corrections as can be seen from (\ref{mninst}). 
There does not seem to be a 4d object that is dual to the D(-1) (see also section 8). 

The field theory instanton in the KS background is a D1-brane wrapped on a 2-cycle of the conifold in order to break the $U(1)_R$. However, a wrapped string will not give the correct real part of the instanton action. The field theory instanton should be a combination of a wrapped string and a D(-1). The motivation for this ansatz comes from (\ref{rgcks}). The sum of the gauge couplings is proportional to the dilaton, which couples to a D(-1), while the difference is proportional to the flux of $B_2$ on the 2-cycle we argued was responsible for the breaking of the $U(1)_R$ symmetry. The action of a D(-1) in this background is constant and real, so D(-1)'s are not responsible for the $U(1)_R$ symmetry breaking. We will again carry out all calculation using the singular background in the UV (the KT background). The value of the induced  RR 2-form over the 2-cycle $\theta=\theta_1=\theta_2, \phi=\phi_1=-\phi_2$ is the same as $B_2$ was in the MN background, $C_{\theta \phi} = \frac{N \psi}{2} \sin \theta$, and therefore will produce the same phase. The real part of the wrapped D1-brane action is  
\be \label{ksinst} S_R = \frac{1}{2\pi \alpha' g_s} \int d\theta d\phi \sqrt{\det (g+2 \pi \alpha' B)} \sim M \sqrt{\ln^2 (u/u_0) + b\ln (u/u_0)}, \ee
where $b$ is some constant. 
In the UV the first term is dominant and we indeed get that the action is proportional to the difference of gauge couplings as in (\ref{rgcks}). Again, there are corrections as we flow to the IR. Note also that neither of the backgrounds reproduces the correct factor of $8 \pi^2$ in (\ref{inst}).

The $Z_{2N}$ non-anomalous R-symmetry
is further spontaneously broken to $Z_2$.
In field theory this breaking of the discrete symmetry is
manifested in the existence of $N$ degenerate vacua. These are the
minima of the superpotential. 
The order parameter associated with this breaking is the gluino
condensate, $\langle \tr \la \la \rangle$.

In the KS and  the MN backgrounds the spontaneous breaking of $Z_{2N}$ down to $Z_2$ is caused by the blow-up of the $S^3$ at the conifold singularity. The existence $N$ degenerate vacua ($M$ in KS) translates in the supergravity picture into the fact that there are only $N$ discrete backgrounds that are truly non-singular \cite{MN}.
Since at the origin the $S^2$ which the F1 or D1 world-sheet wraps is contractable the flux through this cycle has to be an integer multiple of $2\pi$.
Asymptotically, for $\tau \rightarrow \infty$, the flux has this property. One has to find directions along which the $S^2$  can be transported
to the origin, $\tau=0$, without any change of the flux (or a change of the form $2\pi n $). Because the angle $\psi$ is trivially fibered over $S^2$, the directions in question will be along constant $\psi=\psi_0$.
It is clear that $\psi_0=0$ is one such value. The radial $\tau$ component of the 3-form is proportional to $\sin \psi$ so that  along $\psi_0=0$  the wrapped world-sheet can be transported to the origin $\tau=0$ without any change of the flux. One can do the same for any $\psi_0=2 \pi k /N$, $k=0..N-1$. The radial $\tau$ component of the 3-form is now proportional to $\sin(\psi-\psi_0)$, which again vanishes allowing the $S^2$ to be transported to the origin. The $N$ vacua are labeled by the phase of the gluino condensate as we will see in the next section.

\section{The gluino condensate}

In this section we will concentrate on the gluino bilinear, $\tr \la \la (x)$. Most of our arguments will be based on the KS model. It was argued in \cite{KS} that in the UV where $N \gg M$ the anomalous dimensions of operators like $\tr \la \la (x)$ are only of order $O(M/N)$ or less, although the theory is not conformal. If at the bottom of the cascade we are left with pure SYM then the dimension of $\tr \la \la (x)$ is also protected in the IR. 
This is because $\tr \la \la (x)$ is one of a set of operators on the field theory side that are not supposed to get any anomalous dimensions. These
operators are the components of the so-called anomaly multiplet. The
dimensions of the operators in this multiplet are protected by virtue
of the fact that the highest component is the trace of the
energy-momentum tensor, which is a conserved current. The lowest component is the gluino bilinear which has dimension 3.

We would like to associate one polarization of $C_2=C_2^{RR}+iB_2^{NS}$ with
the operator $\tr \la \la(x)$. The value of $G_3=dC_2$ at $\tau \ra \infty$ is (we write only the polarizations along $T^{1,1}$)
\be G_3 \ra \frac{M}{2} (1 + \tau e^{-\tau}) g^5 \wg g^3 \wg g^4 +
\frac{M}{2} (1 - \tau e^{-\tau}) g^5 \wg g^1 \wg g^2 + \frac{ig_sM}{2}
\tau e^{-\tau} g^5 \wg (g^1 \wg g^3 + g^2 \wg g^4) . \ee
If we subtract the asymptotic value of $G_3$, which has nothing to do with chiral symmetry breaking (in fact it is the same as in the KT solution) we get 
\be \Delta G_3 = \frac{M}{2} \tau e^{-\tau} \  \omega_3, \qquad \omega_3=\Big[g^5 \wg (g^3 \wg g^4 -
g^1 \wg g^2) + ig_s g^5 \wg (g^1 \wg g^3 + g^2 \wg g^4) \Big]. \ee
This is the polarization we would like to associate with $\tr \la \la
(x)$. Furthermore, it can be seen from (\ref{gi}) that the corresponding polarization of $C_2$,  
\be C_2 = -\frac{M}{2} \tau e^{-\tau} \ \omega_2, \qquad \omega_2= \Big[(g^1 \wg g^3 + g^2 \wg g^4)
+ ig_s(g^1 \wg g^2 - g^3 \wg g^4)\Big], \ee
 transforms by a phase when $\psi\ra\psi+\delta\psi$, in agreement with fact that $\tr \la \la(x)$ is charged under the broken $U(1)_R$.

For large $\tau$ one can change variables to $u=\epsilon^{2/3}
e^{\tau/3}$. We will use the identification made in \cite{KS} that the deformation parameter  $\epsilon$ is related to the 4d mass scale as $m \sim
\epsilon^{2/3}$. 
Thus, we get that close to the boundary $\Delta G_3$ is of the form
\be \label{g3cl}  \Delta G_3 = \frac{M}{2} \ \frac{m^3}{u^3}
 \ln \frac{u^3}{m^3} \ \omega_3 . \ee 
In a conformal theory (AdS) this is almost the behavior we would expect of a scalar operator of dimension 3 that has a VEV.
We know that in the field theory dual the $Z_{2M}$ symmetry is broken by
the deformation of the conifold to $Z_{2}$. We see from (\ref{g3cl}) that this is in agreement with a non-zero $\llc \sim M m^3$.

The two point correlation function of the gluino bilinear can be
derived by considering the effective action for the polarization of
$G_3$ mentioned above.  
$$ C_2 \ra C_2 + y(x,u) \ \omega_2, $$
\be \label{per} G_3 \ra G_3 + y(x,u)  \ \omega_3+
d y(x,u) \wg  \omega_2, \ee
where $y(x,u)$ is the perturbation with non-vanishing boundary values $y_0(x_1)$ and $y_0(x_2)$. There could and probably are large mixings between these modes and other modes, but this simplified calculation is only intended to show that the 2-point function is space-time independent.

Substituting these perturbations into the relevant part of the supergravity action  
\be \label{sugra} \int d^{4}x du \sqrt{-g} \Big[ G_3 G_3^{*} + \frac{1}{2} \Big( F_5 -
\frac{1}{2i} (C_2 \wg G_3^* -  C_2^* \wg G_3 ) \Big)^2 \Big], \ee
and integration over $u$
will not yield a kinetic term of the schematic form $dy_0(x_1) dy_0(x_2)$. 
The field theory interpretation of this is the well-known result that correlation functions of lowest components of chiral superfields are space-time
independent. It remains to be shown that a mass term of the schematic form $m^6
y_0(x_1)y_0(x_2)$ does originate. It can be seen from (\ref{defcon}) that the supergravity action is indeed proportional to $\epsilon^{12/3} \sim m^6$, through $\sqrt{g}$, and that the quartic term in (\ref{sugra}) will produce a 
term quadratic in $y_0$. 

The two point function of the gluino bilinear with its complex
conjugate, $\langle \tr \la \la (x) \tr {\bar \la} {\bar \la} (y) \rangle$ will receive contributions from glueball exchange \cite{Kra,CH}. Operators like $\tr \la \la (x)$ are only the lowest members of a tower of KK states \cite{CDDF}. One must keep in mind that pure $\N=1$ SYM does not have these operators in its spectrum. If one can show that there are no mixings between these supergravity modes and the lowest mode, then in calculating 2-point functions of point operators one can effectively decouple these KK states. In calculating 2-point functions of extended operators like Wilson loops one has to remove the contributions from such states by hand \cite{LS}, because the string world-sheet can couple to all such operators.

Gluino condensation in the MN background is somewhat more subtle. We can repeat the same analysis that was carried out in the KS background, namely that the difference between the value of $G_3$ in the singular and the deformed solution
is the dual of the gluino bilinear. However, since the UV is dual to some twisted 6d theory we do not have a good understanding of why this is the dual supergravity mode. Another argument, which will be presented in the next section, is that $G_3$ couples to 5-branes, that play the role of domain walls.

\section{BPS Domain walls}

Let us first briefly review the  basic  properties of the domain walls.
$\N=1$ supersymmetric gauge theories, which are characterized  by a discrete set of vacua, generically admit BPS domain wall configurations  that 
interpolate between  the inequivalent vacuum states. 
The domain walls  preserve half of the $\N=1$ supersymmetries.
In the field theory picture 
 such BPS-saturated walls satisfy first-order
differential equations, which follow in a straightforward manner from
the requirement of preserving two out of four global supersymmetries.
Denoting  the  superpotential  and the K\"ahler metric 
by $W(A^i)$  and $g_{j\bar i}(A^i, \bar A^{\bar j})$ respectively,
  where $A^i$ are the chiral superfields 
describing the low energy effective action, 
 and taking the  coordinate 
normal to the wall to be $x_3$, the condition  takes the following form 
\be
\left(Q_\alpha -ie^{i(\Delta W)_{arg}} \sigma^3_{\alpha\dot\alpha}
        {\overline Q}\vphantom{Q}^{\dot\alpha}\right)
{\rm |wall} \rangle =\ 0,
\ee
where $(\Delta W)_{arg}$ is the argument of the difference 
of the superpotential
between the two vacua. This condition translates into the following 
first order differential equation \cite{DvaShi} 
\be
{dA^i\over dx_3}\
=\ e^{i(\Delta W)_{arg}} g^{i\bar j}
        {\partial {W^*}\over \partial \bar A^{\bar j}}.
\ee

The tension of  the  BPS domain wall is
exactly determined by the difference between the superpotential values
in the two vacua connected by the wall.
In the $\N=1$ SYM theory, the superpotential, which acts as a
central charge for domain walls, is related to the gluino condensate, 
so a BPS domain wall has tension \cite{DvaShi}
 \be \T_{DW}\  =\ {N\over 8\pi^2}\,
| \Delta \langle{tr \lambda \lambda}\rangle |. \ee
In the large $N$ limit the domain wall tension is linear in $N$.
The trace yields a factor of $N$ but the phase difference of two vacua is 
proportional to $N^{-1}$. 

In  $\N=1$  theories the effective superpotential
is so constrained by the twin requirements of the holomorphy
and flavor symmetry that one can completely determine
its exact form \cite{VenYan,TVY,AffDinSei}. 
Unfortunately, no such constraints apply to the effective K\"ahler function
of the theory which controls the kinetic energies of the fields.
In fact, due to the lack of knowledge of the latter the story of BPS
domain walls in $\N=1$ is not fully established. 
The K\"ahler function is under much better control in theories  with
$\N=1$ which are derived by mass perturbation of the Seiberg-Witten
 $\N=2$ SYM models \cite{KSY,AKS}.

BPS domain walls were analyzed also in the context of MQCD
\cite{W2,KSY,Vol}. In that formulation
the vacua correspond to various Riemann 2-cycles that the M5 wraps and the 
 BPS-saturated domain walls 
correspond to the supersymmetric 3-cycles that interpolate between those
Riemann surfaces. 
So far analytic expressions for such configurations were not discovered
\cite{KSY}. However, it was shown \cite{W2}  that the parameters of the models
can be chosen in such a way that the string tension does not depend of $N$,
whereas the domain wall tension is linear in $N$.

Recently, domain walls were discussed in  the context of the $\N=1^*$ model 
\cite{PS,BP}. In this model there is a very rich
structure of vacua including both confining and Coulomb phase vacua.
In the former case, namely, configuration that interpolate between
massive vacua the BPS equations again involve the superpotential as
well
as the K\"ahler function expressed in terms of the degrees of freedom
of the low energy effective theory. It is thus clear that those
configurations cannot be determined using the techniques of \cite{BP}

We now turn to the description of the BPS domain walls in the supergravity
duals of theories with  $\N=1$.
Domain walls are attributed
in these models \cite{KS,MN} to D5-branes wrapping the $S^3$.
 Since the volume of $S^3$ is minimized at $\tau=0$ the D5-brane will prefer to be in this region. 
This is in agreement with the fact that only in the IR does the theory have $N$ vacua. 
In the notation of MN the domain wall spans the coordinates $x_0,x_1,x_2,e_i$,($i=3..5$), 
so that it corresponds to turning on a shift of the 3-form $G_3$
which has components along the $S^3$ and depends on $x_3$ in such a way that once one pass from $x_3=-\infty$ to $x_3=\infty$ there is a change of the flux of $B_2$ through $S^2$ in the $S^3$ as was explained in section 4.  
A domain wall interpolating between adjacent vacua corresponds to a single brane since it generates a jump of one unit in the flux whereas $k$ coincident D5-branes are associated with a domain wall connecting vacua that  differ by $k$ units of flux.
Both the KS and the MN backgrounds are of the form $R^7 \times S^3$ near $\tau=0$, with the $S^3$
 having radius proportional to $\sqrt{g_s N}$. Therefore, a D5-brane wrapping the $S^3$ and $x_0, x_1, x_2$ will give a domain wall with tension
\be \T_{DW}= \frac{1}{\alpha'^3 g_s} \int_{S^3} \sqrt{g} = \frac{N}{\alpha'^{3/2}} (g_s N)^{-1/2}. \ee   

The $S^3$ of the conifold is a supersymmetric cycle \cite{GK}. 
It is thus plausible that the
wrapped D5-branes constitute BPS saturated states. 
This can be also examined from the low energy field theory  on the
brane. It was argued recently \cite{AV} that the $\N=2$ (4 supersymmetries) 
inherited from the $\N=1$ in four dimensions is
broken down to $\N=1$, namely two supersymmetries, by  a level $N$ Chern-Simons
term. In the present type IIB setup it is a result of a
$\int C_2\wedge F\wedge F=\int G_3 \wedge CS(A)$ term in the D5
world-volume action, which after integration over the $S^3$ turns into
a CS term at level $N$.
 
Several additional properties of the field theory BPS domain wall can be understood in
the supergravity language:
(i) Crossing a domain wall made out of $k$ wrapped five-branes in the MN model
implies changing the value of $\psi_0$ by
$\Delta \psi_0=2\pi k/N$ which is caused by a change of the flux on the $S^2$ by $k$
units. Since for $k=N$  the value of $\psi_0$ is the same on both sides of the walls it is 
anticipated that in this case the domain wall should fade away. 
Indeed it was shown in \cite{MN} that in the 7d form of the background
the wrapped NS5-brane is charged under the 3-form potential. Due to the
fact that it couples to the $SU(2)_R$ gauge field, the $N$ wrapped five branes can be
replaced by an infinitely large gauge instanton and hence disappear.   

(ii) Since a fundamental string can end on a (wrapped) D5-brane, and since the role of a flux
tube connecting quark anti-quark is played by a fundamental string, it is clear that in the 
supergravity picture of a domain wall flux tubes can end on it.

(iii) Another feature of the domain wall is that a baryon can be dissolve in
it \cite{PS}.
Recall that the baryon vertex is a wrapped D3-brane over the $S^3$ with $N$ strings
attached, so it can be embedded in the wrapped D5-brane so that there are $N$ strings ending
on  the wall. 

\section{Other supergravity brane probes}

Most of the $\N=1$ gauge theory features discussed in the previous sections were associated with certain brane probes. This raises the question of whether all stable brane probe configurations can be attributed to certain properties of the  dual 4d gauge theories. To address this question we summarize in the following table the configurations that have already been discussed together with other possible wrapped and un-wrapped brane probes. We make certain comments about possible interpretation of the latter in the gauge theory picture. 

  \begin{table*}[htbp]
  \begin{center}
\begin{tabular}{|l||c|c|c|c| }
\hline
       & wrapped $S^2$   & wrapped $S^3$ & wrapped $S^3\times S^2$   &un-wrapped  \\
\hline 
\hline 
D(-1)      &  / & /  &  /& instantons, (a)  \\
\hline
F1-strings           &  instantons (a)& / & /&  Wilson loop \\
\hline
D1           &  instantons &  / & /& (b) \\
\hline
D3    & 't Hooft loop   & baryons & /& (c)   \\
\hline 
D5  &  4d  world-volume  & domain walls  & baryons   & / \\
\hline 
$(p,q)$ 5-branes  &  world-volume  & (d)  &  (d)   & / \\
\hline 
D7  &  /  &  (e)  & (e)  & / \\
\hline
\end{tabular}
\end{center}
\end{table*}

(a) As was discussed in section 4,
a D(-1) brane  combined with a wrapped D1-brane
play the role of the field theory instanton in the KS model.
 In the MN background D(-1)'s do not seem to be connected with 4d 
instantons because they do not have the right action. 
It is possible that they correspond to some state in the 6d theory.
Whereas in MN the F1 world-sheet wrapped over $S^2$ is
associated with the instantons in the UV (and the D1 in the IR),
it is not clear to us what is the role of wrapped F1 probes in the KS
model.

(b) Unlike their role in the $N=1^*$ model \cite{PS},
 the un-wrapped D1-branes 
 cannot play the role of the 't Hooft loops connecting
an external monopole anti-monopole pair. This configuration can be thought of as a dimensional reduction of the D5/D1 system (6d instanton) on $S^2$. In the case where the D1 is wrapping the $S^2$ it is a 4d instanton (a). If it does not wrap $S^2$, and we  hold the ends of the D1 on the boundary, then from the gauge theory perspective it is a point-like object with an action linear in $N$.
The gauge theory interpretation of such objects is not clear to us. 
In the MN model a D1-brane in the IR is a F1-string in the UV, where the metric is flat. Therefore, such a string will stay at the boundary. 
It might be identified with a string in the 6d theory.

(c) D3-brane. We start with a configuration of a D3 ending on a D5, and then dimensionally reduce on $S^2$. This configuration is a monopole in the 6d theory. If the D3 is wrapped on $S^2$ it has an interpretation of a 4d monopole. If the D3 does not wrap $S^2$, and we hold one of its ends on the boundary, it looks like an infinite tension (external) domain wall from the 4d perspective \cite{GRW}.

(d) $(p,q)$-strings and 5-branes.
In $N=1^*$ theory \cite{PS} the un-wrapped $(p,q)$ strings serve as the probes of the $(p,q)$ vacua \cite{PS,KLSSY}. Since this theory is a daughter theory of the $SL(2,Z)$  invariant $\N=4$ theory  these strings can be obtained
from F1-strings by an $SL(2,Z)$ transformation. 
In $\N=1$ SYM there are no remnants of the  $SL(2,Z)$
structure 
and indeed the $(p,q)$ vacua do not exist, so it is not clear  
what is the interpretation of $(p,q)$ strings both
un-wrapped and those that wrap the $S^2$. 
If one does not find any reason to dismiss this kind
of probes in the supergravity picture, it may hint that the vacua of
the $\N=1$ holographic dual of the supergravity backgrounds discussed, are 
characterized by additional
order parameters apart from the gluino condensate. Recall that in 
 the perturbed $\N=2$ Seiberg-Witten theory in addition to the expectation values $U=\langle Tr \Phi^2 \rangle$ there are also monopoles (or dyons) condensates \cite{KSY,AKS}. 
It might be that the instanton like objects,
associated with the wrapping of the $(p,q)$ strings, are related to these
additional order parameters. 
In a similar manner, it might be that the wrapping of the $(p,q)$
5-branes over the $S^3$ are
associated with the domain walls profiles of the various monopole and dyon fields \cite{KSY,AKS} that interpolate between the values of these fields at the different vacua. 
One probably has to  treat the ``baryons'' associated with the wrapping of $(p,q)$ 5-branes over the $S^3 \times S^2$ on the same footing.

(e) Wrapped D7-brane over the $S^3$ are points on the $S^2$ and
as such do  not have 4d duals. By wrapping D7-branes over $S^2\times S^3$ we get a  membrane from the 4d point of view. Again one may want to interpret them as domain walls but together with the wrapped D5 branes and the unwrapped D3 branes it is clear that supergravity offers too rich a
set of domain wall candidates, and presumably only the former has the appropriate 4d field theory interpretation.


\section{ Guide lines for constructing 
  supergravity duals of $\N=1$ gauge dynamics}

So far we have investigated the realization of  properties of  $\N=1$ gauge  theories  
mainly in the backgrounds of KS and MN. 
We now would like to examine how such  properties constrain
the construction of any supergravity background dual to a confining
$\N=1$ gauge  theory, such as the general class proposed in \cite{FS}.
In particular we analyze  the following properties:  
(i) Gauge group,
(ii) Supersymmetries, (iii) Wilson loops  ('t Hooft loops)  and the corresponding   
 quark anti-quark  (monopole anti-monopole) potential, (iv) Instantons and the $U(1)_R \ra Z_{2N}$ symmetry breaking, (v) gluino condensation and the spontaneous $Z_{2N} \ra Z_2$ breaking,  (vi) monopoles, (vii) domain walls, (viii) baryons and (ix) KK states.  
This discussion does not constitute a set of
restrictive rules but rather certain insights based on the experience gained from the analysis of existing models.   
 
\begin{itemize}

\item 
In the original AdS/CFT duality the number of colors $N$ in the
boundary $SU(N)$ gauge theory is associated with the number of the D3-branes or the flux of the 5-form.
To have $SO(N)$ or $SP(N/2)$ gauge groups required an orientifold
operation that replaces the $S^5$ with an $RP^5=S^5/
  Z_2$  \cite{FaySpa}. 
In orbifold models one mods with a discrete group, $\Gamma\in
SU(4)$. If the set of irreducible representation of $\Gamma$ is $\{
r_i\}$ with corresponding dimension $d_i$, then starting with a 
gauge group $U(\sum_i d_i N)$ one finds after the orbifolding
$\prod_i U(d_i N)$. 
For an Abelian discrete group the gauge 
group is $ U(N)^{|\Gamma|}$ where
$|\Gamma|$ is the order of the discrete group $\Gamma$
and the gauge group before orbifolding is $U(\Gamma N)$.
In particular for $\Gamma=Z_k$ one gets $U(N)^k$.
Another mechanism to generate holographic gauge groups 
is to consider supergravity backgrounds based placing D3-branes
on conifold singularities. The original model based on
$T^{1,1}$ \cite{KW} was generalized to ADE conifolds \cite{GNS}.  
The gauge theories dual to both the orbifolds and the conifolds
include matter superfields in bi-fundamental  representations.
To build a supergravity dual to a theory with a single $SU(N)$ group 
factor (in $\N=1$), wrapped 5-branes seem to be useful either
following the lines of the MN model, or as a result of 
a Seiberg duality cascade as in the KS model.
To get a single $SO(N)$ or $SP(N/2)$ groups an orientifold will be 
 needed.

\item
At least four  mechanisms for breaking the maximal 16 supersymmetries 
down to the 4 supersymmetries of $\N=1$ were suggested:
(i) Orbifold models constructed from D3-branes at $R^6/ \Gamma$ orbifold
singularities where $\Gamma\in SU(3)$ \cite{KasSil}
(ii) Conifold models based on D3-branes at 
the singular point of a conifold with a base like the 
 $T^{1,1}$  \cite{KW, KS}
(iii) Wrapping the little string theory on $S^2$ and   
embedding the spin connection of the $S^2$ in a particular way in the $SO(4)$ isometry
group of the NS5-brane solution \cite{MN}.  
(iv) A soft breaking via a Myers mechanism \cite{RM} due to incorporating a RR magnetic 3-form in the AdS background \cite{PS,GP,Gub1}. 
 For duals of superconformal gauge theories (i) and (ii) 
may serve as a starting point. To construct supergravity duals of field 
theories with soft breaking down to $\N=1$ one can use (iv).
For theories that in the IR resemble the $\N=1$ SYM
 fractional branes should be added to the orbifold and conifold
models or one can use the twisting approach of (iii). 
In fact it turns out that the two approaches (ii) and (iii) are
closely related. It is easily shown that like in the MN model,
in the conifold model of \cite{KW}, the spin connection along the
$S^2$ is identified with the $U(1)_R \in SU(2)_R$.
Further more, an interpolation between the KS model and the wrapped
D5-brane solution of MN was written down in \cite{TP}. 
These cases were shown to be special cases of an ansatz that
describes spaces with topology  
$R\times S^2\times S^3$.
The general conditions that a background of the form $R^{1,3}\times
M^6$ has to be obey in order to  preserve four supersymmetries are \cite{Str,PV,TP}
(i) $M^6$ should be a Hermitian manifold, (ii) the connection with
torsion  should have its holonomy contained in $SU(3)$, 
(iii) The K\"ahler form and the complex structure should obey 
the dilatino Killing spinor equation. 
To construct $\N=1$ models one may use one of the conifolds of Calabi-Yau 
compactification, or the method of fractional branes for instance with 5-branes wrapping two cycles other than the $S^2$.

\item
Models dual to a gauge theory in the confining phase have to admit Wilson 
loops with area law behavior \cite{Cobi}. 
As discussed in section 2, for cases
with no additional 5-branes,  this implies that  either
(i) $f(\tau_{min})>0$ or that  (ii) $f(\tau_{div})>0$ \cite{KSS1}. 
In models with additional 5-branes as part of the background,  
 the world-volume physics on these branes should also be invoked \cite{KLSSY}.
Since F1-strings cannot end on NS5 branes, but D1-branes can, localizing   
such branes as part of the background may induce confinement.  
For models without such branes 
the question is what option out of the two (with $\tau_{min}$ or with $\tau_{div})$ is  preferable.
If in option (i) $\tau_{min}$ is not the minimal value of $\tau$
then there might be an additional 4d boundary at a lower value of $\tau$.
In that case one has to check that
signals cannot propagate between the two boundaries \cite{AFS}. 
On the other hand option (ii), that is associated with a horizon, may result in a divergent string tension \cite{BISY}.
As was explained in section 2, any generic solution for which the string that corresponds to the Wilson loop is mostly a flat string along  $\tau_{min}$ (or $\tau_{div})$ ), admits
the desired features of the flux tube similar to those  derived  in flat space time.
The counterpart of a confining Wilson loop should be a screening 
't Hooft loop. 
Replacing the F1 connecting the external quark anti-quark pair with 
a D1 attached to a monopole anti-monopole system (or wrapped D2 in IIA) properly describes a 't Hooft loop in certain models \cite{BISY,Li,GrosOo}. 
Localized NS5-branes which are part of the background will induce screening behavior since the D1-branes can end on them.
For backgrounds without such branes,
since the action of the D1 strings is related to that of the ordinary 
strings by $e^{-\phi}$, such a construction can produce screening
only provided that 
$e^{-\phi(\tau_{min})}$ vanishes (or similarly with $\tau_{div}$).
 This requires that the string coupling diverge at
some point along the radial coordinate and hence it cannot serve as the supergravity dual
mechanism of screening of monopoles. In type IIB performing S duality is not 
helpful here since it  will transform the system to that of an F1.
Instead one can 
attribute the 't Hooft loop to a fractional D1 that corresponds to 
a wrapped Dp-brane over a compact $p-1$ cycle.
This mechanism is viable provided that the 
volume of the cycle
vanishes at some value of the radial coordinate. 

\item
To account for the gauge theory instantons, one may use D(-1)  
brane probes or wrapped Euclidean Dp-branes on $p+1$ cycles, or some
combination of the two. From the relation to
the gauge coupling, the 
breaking of the $U(1)_R$ symmetry and the existence of $N$ degenerate vacua,
in the MN and KS models, one can deduce a supergravity scenario of
$\N=1$ instantons that includes the following ingredients:
(i) The imaginary part of the action of the wrapped branes 
 is proportional to $N$ times the  angle whose
shift transformation corresponds to the $U(1)_R$. In this way, the 
requirement that the change of the action under the symmetry
transformation is a multiple of $2\pi$, restricts the shifts only to
elements of $Z_{2N}$. This means that 
 the angle discussed has to be trivially
fibered over the cycle the brane is wrapping.
(ii) The real part of the action should be proportional to $g_{YM}^{-2}$.
(iii)  The cycle which is wrapped  collapses to zero  in  
the region  that corresponds to the IR. 
Since the flux on a collapsing cycle has to be multiple of
$2\pi$, non-singular solutions can be achieved when the flux in the UV
is a multiple of $2\pi$ and there is no change in the flux as a
function of the radial direction.  Combined with the previous
ingredient that the flux is linear in the angle, and $N$, this 
requirement selects a set of only $N$ values of the
angle and hence $N$ degenerate vacua.
For instance one may imagine D0-branes of type IIA with Euclidean time
direction that wrap an $S^1$. The real part of the action is proportional  
 to the circumference of the circle. The imaginary part to $N$ times
 an angle that is perpendicular to the $S^1$. The radius of the $S^1$
 should vanish in the IR.

\item
In $\N=1$ models, in which gluino condensation is expected, there
should be some complex supergravity mode (in general a mix of modes)
that approaches the boundary in such a way that implies  that the dual
operator has a non-zero VEV. In the KS model this combination of modes
reduces in the UV to one of the polarizations of $C_2$. In a general
background it is impossible to guess what the dual supergravity mode
will be, and one must calculate on a case by case basis. In \cite{Bu}
finite temperature generalizations of the KS and MN
backgrounds were found. In these solutions there is no chiral symmetry
breaking and therefore no gluino condensate.

\item 
A probe configuration that can be associated with the $\N=1$ domain
wall is a wrapped Dp-brane on a $p-2$ cycle ($p>3$) that admits BPS
solutions. If the background is a function of $g_s N$ (and not of 
$N$ separately, then in the large $N$ limit 
the domain wall tension is necessarily linear in $N$,
since the action of a Dp-brane is proportional to $g_s^{-1}=N \la^{-1}$. 
Similar to the construction in MN and KS the interpolation between two
vacua that differ by a phase of ${2\pi i k\over N}$ is associated with
a group of $k$ wrapped branes. The dual of the flux tube,
an F1 string connecting two points on the boundary, can end on a
single wrapped Dp-brane; whereas on a stack of $k$ such branes only 
$k$ coincident strings can end. This is due to the fact there is
a 3d $U(k)$ gauge theory on the $k$ domain walls and the end of a
string on this stack of branes is a quark in the fundamental
representation. In fact, if the 3d  $\N=1$  $U(k)$ gauge theory
has a Chern-Simons term \cite{AV}, the above statement has
to be reexamined.

\item
The role of baryons can be played by wrapping of Dp-branes over p-cycles. Baryon vertex  connected to $N$ external fundamental quarks is dual to such a wrapped brane which is necessarily connected to $N$ strings to conserve the charge associated with the $N$ units of flux. 
As was shown in section 4 such baryons have mass which is $N$ times that of the corresponding
``mesons''.   
Baryons composites of elementary particles that are incorporated in the background (as oppose
to external ones) can be constructed from wrapped branes connected with strings between themselves and not with the boundary. In particular if on top of the wrapped Dp-brane
there are also Dp'-branes on $p'$ cycles then one can
form objects made out of $k$ wrapped Dp-branes connected with $kN_p=k'N_{p'}$ strings
to $k'$ wrapped Dp'-branes.
 Since in $\N=1$ gauge theory the baryons cannot be BPS states, the wrapped branes should also have no left over supersymmetries.

\item 
A common feature to all the supergravity backgrounds dual to $\N=1$ gauge
theories is the KK modes that have masses of the same order
of magnitude as the glueballs. To  decouple these states one needs to be in the region where $\alpha' {\cal R} \sim (g_sN)^{-1/2} \gg 1$,
namely, in a region where the curvature is large and the supergravity approximation is not valid.  
The question is whether one can get rid of these modes without
destroying the desirable features of the models.
One alternative to avoid the excitations associated with the $S^2$
and $S^3$ cycles is to have a non-critical model without the extra
compactified dimension. A possible mechanism for that may be the
replacement of the conifold by some CFT model like the 
$c=1$ super Liouville model \cite{Vafa} that should obviously also
incorporate RR fields.  The problem with this approach is two folded: (i) to
write down the appropriate CFT model (ii) to find a way to mimic all
the non-perturbative properties associated with the wrapping of branes
over these cycles.
Another alternative is to have two scales in the problem.
One big mass scale associated with the compact cycle which is the host 
of the wrapped branes, and another scale corresponding to the QCD scale
which is not characterized by a compact cycle but rather by a distance between
non compact branes like in MQCD. 

\end{itemize}   

{\bf Acknowledgments}  We would like to thank Yaron Kinar, Ehud Schreiber and Shimon Yankielowicz for useful discussions. We would especially like to thank Ofer Aharony for numerous discussions and a critical reading of the manuscript,
 and Igor Klebanov for his comments.
This work was  supported in part 
by the US-Israel Binational Science
Foundation, by GIF -- the German-Israeli Foundation for Scientific Research,
and by the Israel Science Foundation. 


\end{document}